\documentclass[twoside]{IEEEtran}
\usepackage{multirow}
\usepackage{subfigure}
\newcommand{\vect}[1]{\boldsymbol{#1}}  
\usepackage{epsfig}
\newtheorem{lemma}{Lemma}
\newtheorem{theorem}{Theorem}

\usepackage{cite}
\usepackage[cmex10]{amsmath}

\usepackage{algorithmic}

\newsavebox{\ieeealgbox}
\newenvironment{boxedalgorithmic}
  {\begin{lrbox}{\ieeealgbox}
   \begin{minipage}{\dimexpr\columnwidth-2\fboxsep-2\fboxrule}
   \begin{algorithmic}}
  {\end{algorithmic}
   \end{minipage}
   \end{lrbox}\noindent\fbox{\usebox{\ieeealgbox}}}

\begin{document}
%
\title{The Entropy of Conditional Markov Trajectories}

\author{Mohamed~Kafsi,~\IEEEmembership{Student Member,~IEEE,}
        Matthias~Grossglauser,~\IEEEmembership{Member,~IEEE,}
        and~Patrick~Thiran,~\IEEEmembership{Senior Member,~IEEE}
\thanks{The authors are with the School of IC, EPFL, Lausanne CH-1015,
Switzerland. Copyright (c) 2013 IEEE. Personal use of this material is permitted.  
However, permission to use this material for any other purposes must be 
obtained from the IEEE by sending a request to pubs-permissions@ieee.org.}}

\markboth{IEEE TRANSACTIONS ON INFORMATION THEORY}%
{The Entropy of Conditional Markov Trajectories}

\maketitle

\begin{abstract}
To quantify the randomness of Markov trajectories with fixed initial and final states, Ekroot and Cover proposed a closed-form expression for the entropy of trajectories of an irreducible finite state Markov chain. 
Numerous applications, including the study of random walks on graphs, require the computation of the entropy of Markov trajectories conditional on a set of intermediate states. 
However, the expression of Ekroot and Cover does not allow for computing this quantity. 
In this paper, we propose a method to compute the entropy of conditional Markov trajectories through a transformation of the original Markov chain into a Markov chain that exhibits the desired conditional distribution of trajectories. Moreover, we express the entropy of Markov trajectories\textemdash a global quantity\textemdash as a linear combination of \textit{local entropies} associated with the Markov chain states.  
\end{abstract}


\IEEEPARstart{Q}{uantifying} the randomness of Markov trajectories has applications in graph theory~\cite{FamilyDiss} and in statistical physics~\cite{Lloyd:Complex}, as well as in the study of random walks on graphs~\cite{MERW,RSP}. 
The need to quantify the randomness of Markov trajectories first arose when Lloyd and Pagels~\cite{Lloyd:Complex} defined a measure of complexity for the macroscopic states of physical systems. 
They examine some intuitive properties that a measure of complexity should have and propose a universal measure called \textit{depth}. 
They suggest that the depth of a state should depend on the complexity of the process by which that state arose, and prove that it must be proportional to the Shannon entropy of the set of trajectories leading to that state. 
Subsequently, Ekroot and Cover~\cite{EntropMarkTraj} studied the computational aspect of the depth measure. 
In order to quantify the number of bits of randomness in a Markov trajectory, they propose a closed-form expression for the entropy of trajectories of an irreducible finite state Markov chain. 
Their expression does not allow, however, for computing the entropy of Markov trajectories conditional on the realisation of a set of intermediate states. 
Computing the conditional entropy of Markov trajectories turns out to be very challenging yet useful in numerous domains, including the study of 
mobility predictability and its dependence on location side information.  

Consider a scenario where we are interested in quantifying the predictability of route-choice behaviour. We can model the mobility of a traveller as a weighted random walk on a graph whose vertices represent locations and edges represent possible transitions~\cite{RouteChoice}. We can therefore model a route as a sample path or trajectory in a Markov chain. 
If we suppose that we know where the traveller starts and ends her/his route, the randomness of the route she/he would follow is represented by the distribution of trajectories between the source and destination vertices. Consequently, the predictability of her/his route is captured by the entropy of Markov trajectories between these two states.  
Now, if we obtain side information stating that the traveller went (or has to go) through a set of intermediate vertices, quantifying the evolution of her/his route predictability requires the computation of the trajectory entropy conditional on the set of known intermediate states. 
The conditional entropy is also a way to quantify the informational value of the intermediate states revealed. 
For example, if the entropy conditional on the set of known intermediate states is zero, then this set reveals the whole trajectory of the traveller.  

In our work, we propose a method to compute the entropy of Markov trajectories conditional on a set of intermediate states. The method is based on a transformation of the original Markov chain so that the transformed Markov chain exhibits the desired conditional distribution of trajectories. 
We also derive an expression that enables us to compute the entropy of Markov trajectories, under conditions weaker than those assumed in~\cite{EntropMarkTraj}. Moreover, this expression links the entropy of Markov trajectories to the local entropies at the Markov chain states.

\section{The model}
\label{Sec:Model}
Let $\lbrace X_{i} \rbrace$ be a finite state irreducible and aperiodic Markov chain (MC) 
with transition probability matrix $P$ whose elements are the transition probabilities 
\begin{align*}
P_{x_{n}x_{n+1}} &= p(X_{n+1} = x_{n+1}|X_{n} = x_{n}) \\
&=  p(X_{n+1} = x_{n+1}|X_{n} = x_{n},\ldots ,X_{1} = x_{1}).
\end{align*}

This MC admits a stationary distribution $\Pi$, which is the unique solution of  
$\Pi = \Pi \, P.$
The entropy rate $H(X)$ is a measure of the average entropy growth of a sequence generated
by the process $\lbrace X_{i} \rbrace$ and is defined as 
\begin{equation*}
H(X) = \lim_{n\rightarrow \infty} \frac{1}{n} H(X_1,X_2,...,X_n).
\end{equation*}
For the particular case of an irreducible and aperiodic MC, the limit above is equal to \cite[p.~77]{EltsInfoTheory}
\begin{equation*}
\label{eq:entropy_rate}
H(X) = \sum_{i} \Pi(i) H(P_{i \cdot}),
\end{equation*} 
where $P_{i.}$ denotes the $i^{\text{th}}$ row of $P$ and where
$H(P_{i \cdot}) = -\sum_{j} P_{ij} \log(P_{ij})$ is the \emph{local entropy} 
of state $i$.
Note that, throughout this paper, we use $MC_P$ as a shorthand for the Markov chain whose transition probability 
matrix is $P$. 
\subsection{The Entropy of Markov Trajectories}
\label{Sec:entrMarkTraj}
We follow the setting of~\cite{EntropMarkTraj} closely. We define a \emph{random trajectory} $T_{sd}$ of a 
MC as a path with initial state $s$, final state $d$, and no intermediate state $d$, i.e., the trajectory is terminated as soon as it reaches state $d$.
Using the Markov property, we express the probability of a particular trajectory 
$t_{sd} =sx_2...x_kd$ given that $X_1 = s$ as 
\begin{equation*}
p(t_{sd}) = P_{sx_2}P_{x_2x_3} \ldots P_{x_kd}.
\end{equation*}
Let $\mathcal{T}_{sd}$ be the set of all trajectories that start at state 
$s$ and end as soon as they reach state $d$. As the MC defined by the matrix $P$ is finite and irreducible,
we have 
\begin{equation*}
\sum_{t_{sd} \in \mathcal{T}_{sd}} p(t_{sd}) = 1\qquad \text{ for all } s,d.
\end{equation*}
So the discrete random variable $T_{sd}$ has as support the set $\mathcal{T}_{sd}$,  
with the probability mass function $p(t_{sd})$. Subsequently, we use $p(t_{sd})$ as a 
shorthand for $p(T_{sd}=t_{sd})$. We can now express the entropy of the random trajectory
${T}_{sd}$ as
\begin{equation*}
\label{eq:Entropy}
H_{sd} \equiv  H({T}_{sd}) = - \sum_{t_{sd} \in \mathcal{T}_{sd}} p(t_{sd}) \log p(t_{sd}).
\end{equation*}
We define the matrix of trajectory entropies $H$ where $H_{ij} = H(T_{ij})$.
Ekroot and Cover~\cite{EntropMarkTraj} provide a general closed-form expression 
for the matrix $H$ of an irreducible, aperiodic and finite state MC.

The entropy $H_{sd \vert u}$ of a trajectory from $s$ to $d$ given that it goes through $u$ is defined by 
\begin{align}
\label{eq:conditionalEntropy}
H_{sd \vert u} & \equiv  H(T_{sd}\vert T_{sd} \in \mathcal{T}_{sd}^u) \nonumber \\
&=  -\sum_{t_{sd} \in \mathcal{T}_{sd}^{u}} p(t_{sd} \vert T_{sd} \in \mathcal{T}_{sd}^{u}) \log p (t_{sd} \vert T_{sd} \in \mathcal{T}_{sd}^{u}),
\end{align} where $\mathcal{T}_{sd}^{u}$ is the set of all trajectories
in $\mathcal{T}_{sd}$ with an intermediate state $u$
\begin{equation*}
\mathcal{T}_{sd}^{u} = \lbrace t_{sd} \in \mathcal{T}_{sd}: t_{sd} = s\ldots u \ldots d \rbrace .
\end{equation*}  
The major challenge is to compute efficiently the entropy $H_{sd \vert u}$. Even the costly approach 
of computing all the terms of the sum~($\ref{eq:conditionalEntropy}$) is not always possible because the set 
$\mathcal{T}_{sd}^u$ has an infinite number of members in the case where, after removing state $d$, the transition graph of the MC
is not a DAG. It is important to emphasize that the entropy $H_{sd \vert u}$ is not the entropy of the random variable $T_{sd}$ given another random variable\textemdash a quantity which is easy to compute\textemdash but the entropy of $T_{sd}$ conditional on the realization of a dependent random variable. 

\begin{figure}[hdt]
\centering
\epsfig{file=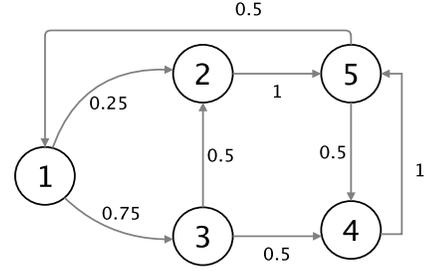,height=1.5in, width=2.25in}
\caption{An irreducible, 5-state, Markov chain annotated with the transition probabilities.}
\label{fig:MCexample}
\end{figure}
In Figure~\ref{fig:MCexample}, we show an example of a finite-state irreducible and aperiodic MC. 
Note that the presence of cycles implies that the set of trajectories between some pair of states 
might have infinite cardinality ($\left|{\mathcal{T}_{14}}\right| = \infty$ for example). Therefore, in 
addition to being complex, the naive approach of enumerating all trajectories is not always possible.

Using the results of \cite{EntropMarkTraj}, we obtain the matrix of trajectory entropies 
\begin{equation*}
H = 
\begin{pmatrix}
3.56&3.69&1.74&3.18&1.56\\
2&5.69&3.74&2.59&0\\
3&3.84&4.74&2.29&1 \\
2&5.69&3.74&2.59&0 \\
2&5.69&3.74&2.59&1.78
\end{pmatrix}.
\end{equation*}
The zero elements of the matrix $H$ correspond to deterministic trajectories
such as $T_{25}$, which is equal to the path $25$ with probability $1$ since no 
other path allows a walk to go from $2$ to $5$. The entropy of the random trajectory
$T_{15}$ is 1.56 bits. Now imagine that we have an additional piece of information stating that the trajectory $T_{15}$ goes through state $4$. Intuitively, we 
would be tempted to argue that the entropy $H_{15 \vert 4}$ of the trajectory $T_{15}$ 
conditional on going through state $4$ is equal to $H_{14}+H_{45}$, but this additivity 
property does not hold. Indeed, the conditional entropy $H_{15 \vert 4}$ is zero because 
the trajectory $T_{15}$, conditional on the intermediate state 4, can only be equal to the path $1345$, whereas $H_{14} = 3.18$ bits, 
hence $H_{14}+H_{45} = 3.18 + 0 = 3.18 \neq H_{15 \vert 4} $ bits.

In the next section, we study the entropy of Markov trajectories 
conditional on \emph{multiple} intermediate states and derive a general expression for this entropy.

\section{The Entropy of Conditional Markov Trajectories}
\label{Sec:EntropCondMarkTraj}
Let $\alpha_{sud}$ denote the probability that the random trajectory $T_{sd}$ goes through the 
state $u$ at least once:
\begin{equation*}
\alpha_{sud} = p(T_{sd} \in \mathcal{T}_{sd}^{u}).
\end{equation*}
This is also equal to the probability that a walk reaches the state $u$ before the state $d$, given that it started 
at $s$. In order to compute $\alpha_{sud}$, the technique from~\cite{Stewart:PMCQS,snell:FiniteMC} 
is to make the states $u$ and $d$ absorbing (a state $i$ is absorbing if and only if $P_{ii} = 1$) and compute 
the probability to be absorbed by state $u$ given that the trajectory has started at state $s$.

Our first step towards computing $H_{sd \vert u}$ is to express it as a function of quantities that are much simpler to compute. 
The idea is to relate the entropy of a trajectory conditional on a given state to its entropy conditional on \emph{not} going through that state. 
Therefore, we define the entropy $H_{sd \vert \bar{u}}$ of a trajectory from $s$ to $d$ given that it does \emph{not} go through $u$ to be
\begin{equation*}
H_{sd \vert \bar{u}} \equiv H(T_{sd}\vert T_{sd} \notin \mathcal{T}_{sd}^u)
.
\end{equation*}
Using the chain rule for entropy, we can derive the following equation which 
relates $H_{sd \vert u}$ to $H_{sd}, H_{sd \vert \bar{u}}$ and $\alpha_{sud}$:
\begin{equation}
\label{eq:ChainRuleEntropy}
H_{sd} = \alpha_{sud} H_{sd \vert u} + (1-\alpha_{sud}) H_{sd \vert \bar{u}} + h(\alpha_{sud})
\end{equation} for all $u$, where $h(\alpha_{sud})$ is the entropy of a Bernoulli random variable with success probability 
$\alpha_{sud}$.  
\begin{proof}
First, we define the indicator variable $I$ by
\begin{equation*}
I  = 
\begin{cases}
1 & \text{if } T_{sd} \in \mathcal{T}_{sd}^{u},\\
0 & \text{otherwise}.
\end{cases}
\end{equation*}
Using the chain rule for entropy, we express the joint entropy $H(T_{sd},I)$ in two different ways, 
\begin{align*}
H(T_{sd},I)  &=  H(I) + H(T_{sd}\vert I) \\
&=  H(T_{sd}) + H(I \vert T_{sd}) = H(T_{sd}),
\end{align*}
because $I$ is a deterministic function of $T_{sd}$. So the entropy of the random trajectory $T_{sd}$ can be expressed as
\begin{align*}
H(T_{sd}) & =  H(I) + H(T_{sd}\vert I)\\
& =  H(I) + H(T_{sd}\vert I = 1) p(I=1)\\ 
& \quad +  H(T_{sd}\vert I = 0) p(I=0)\\
& =  H(I) + H(T_{sd}\vert T_{sd} \in \mathcal{T}_{sd}^u) p(T_{sd} \in \mathcal{T}_{sd}^u) \\
& \quad +  H(T_{sd}\vert T_{sd} \notin \mathcal{T}_{sd}^u) p(T_{sd} \notin \mathcal{T}_{sd}^u).
\end{align*}
Since $\alpha_{sud} = p(T_{sd} \in \mathcal{T}_{sd}^{u}) = p(I=1)$, we obtain 
\begin{align*}
H(T_{sd}) &=  \alpha_{sud} H(T_{sd}\vert T_{sd} \in \mathcal{T}_{sd}^u) \\
& \quad +   (1-\alpha_{sud}) H(T_{sd}\vert T_{sd} \notin \mathcal{T}_{sd}^u) + h(\alpha_{sud}).
\end{align*}
\end{proof}
As we know from~\cite{EntropMarkTraj, Stewart:PMCQS, snell:FiniteMC} how to compute $H_{sd}$ and $\alpha_{sud}$, if we are able to compute $H_{sd \vert \bar{u}}$,
we can use~(\ref{eq:ChainRuleEntropy}) to find $H_{sd \vert u}$. 
However, generalizing~(\ref{eq:ChainRuleEntropy}) to trajectories conditional on passing through
\emph{multiple} intermediate states turns out to be difficult, hence we propose an approach that circumvents this problem. As we will see, the difficulty of our approach also boils down to computing the entropy of a trajectory conditional on \emph{not} going through a given state. 

First, we define $\mathcal{T}_{sd}^{\vect{u}}$, the set of all trajectories in $\mathcal{T}_{sd}$ that 
exhibit the sequence of intermediate states $\vect{u} = u_1 u_2 \ldots u_l$, i.e. 
\begin{equation*}
\mathcal{T}_{sd}^{\vect{u}} = \lbrace t_{sd} \in \mathcal{T}_{sd}: t_{sd} = s\ldots u_1\ldots u_2\ldots u_l\ldots d \rbrace .
\end{equation*}  

For an arbitrary sequence of states $\vect{u} = u_1 u_2 \ldots u_l$, satisfying $p(T_{sd} \in \mathcal{T}_{sd}^{\vect{u}}) > 0$, we prove the following lemma. 
\vspace*{0.25cm}
\begin{lemma}
\label{thm:cond_entropy_as_sum}
\begin{equation}
\label{eq:CondiEntropy}
H(T_{sd}\vert T_{sd} \in \mathcal{T}_{sd}^{\vect{u}}) = \sum_{k=0}^{l-1} H_{u_ku_{k+1} \vert \bar{d}} + H_{u_ld},
\end{equation} 
where $u_0 = s$.
\end{lemma}
\vspace*{0.25cm}

\begin{proof}
First, given $T_{sd} \in \mathcal{T}_{sd}^{\vect{u}}$, the random trajectory $T_{sd}$ can be expressed as a sequence of random sub-trajectories $(T_{su_1},T_{u_1u_2},\ldots, T_{u_{l-1}u_l},T_{u_ld})$. 
Therefore, the conditional entropy $H(T_{sd}\vert T_{sd} \in \mathcal{T}_{sd}^{\vect{u}})$, which we denote by $H_{sd\vert u_1\ldots u_l}$, can be written as a joint sub-trajectory entropy
\begin{equation*}
H_{sd\vert u_1\ldots u_l} = H(T_{su_1},T_{u_1u_2},\ldots,T_{u_ld}\vert T_{sd} \in \mathcal{T}_{sd}^{\vect{u}}).
\end{equation*}
Applying the chain rule for entropy, we obtain successively
\begin{align*}
H_{sd\vert u_1\ldots u_l} & =  H(T_{su_1},T_{u_1u_2},\ldots ,T_{u_ld}\vert T_{sd} \in \mathcal{T}_{sd}^{\vect{u}}) \\
& =  H(T_{su_1} \vert T_{sd} \in \mathcal{T}_{sd}^{\vect{u}}) \\ 
& \quad + H(T_{u_1u_2} \vert T_{su_1};T_{sd} \in \mathcal{T}_{sd}^{\vect{u}}) \\
& \qquad \vdots \\
&\quad + H(T_{u_ld}\vert T_{su_1},\ldots,T_{u_{l-1}u_l};T_{sd} \in \mathcal{T}_{sd}^{\vect{u}}).
\end{align*}
The Markovian nature of the process generating the trajectory $T_{sd}$ implies that each of
the sub-trajectories $T_{u_ku_{k+1}}$ is independent of the preceding ones, given its starting point $u_k$. 
Since the sequence $s \vect{u} = s u_1 u_2 \ldots u_l $ defines the starting point of each sub-trajectory, we can 
therefore write that  
\begin{align}
\label{eq:indep}
&H(T_{u_ku_{k+1}}\vert T_{su_1},\ldots,T_{u_{k-1}u_k};T_{sd} \in \mathcal{T}_{sd}^{\vect{u}}) \nonumber \\
&= H(T_{u_ku_{k+1}}\vert T_{sd} \in \mathcal{T}_{sd}^{\vect{u}}).
\end{align}
Using~\eqref{eq:indep}, the expression for the conditional entropy becomes  
\begin{align*}
H_{sd\vert u_1\ldots u_l} &= H(T_{su_1}\vert T_{sd} \in \mathcal{T}_{sd}^{\vect{u}}) \\ 
&\quad + H(T_{u_1u_2}\vert T_{sd} \in \mathcal{T}_{sd}^{\vect{u}}) \\
& \qquad \vdots  \\ 
&\quad + H(T_{u_ld}\vert T_{sd} \in \mathcal{T}_{sd}^{\vect{u}}).
\end{align*}
Note that for each trajectory $T_{u_ku_{k+1}}$, the only restriction imposed by the event 
$\left\lbrace T_{sd} \in \mathcal{T}_{sd}^{\vect{u}} \right\rbrace $ 
is that the final state $d$ cannot be an intermediate state of any of the first $l$ trajectories $T_{su_1},T_{u_1u_2},\ldots, T_{u_{l-1}u_l}$. 
As a result,
\begin{align*}
H_{sd\vert u_1\ldots u_l} &= H(T_{su_1}\vert T_{su_1} \notin \mathcal{T}_{su_1}^{d}) \\
& \quad + H(T_{u_1u_2}\vert T_{u_1u_2} \notin \mathcal{T}_{u_1u_2}^{d})\\
& \qquad \vdots \\
&\quad + H(T_{u_ld}) \\
&=  \sum_{k=0}^{l-1} H_{u_ku_{k+1} \vert \bar{d}}+ H_{u_ld},
\end{align*}
where $u_0=s$
\end{proof}
Now, if we are able to compute $H_{u_ku_{k+1} \vert \bar{d}}$, we can use~(\ref{eq:CondiEntropy}) to derive $H(T_{sd}\vert T_{sd} \in \mathcal{T}_{sd}^{\vect{u}})$. 
The following lemma shows how the conditional entropy $H_{u_ku_{k+1} \vert \bar{d}}$ can be obtained by a simple modification of the MC.\\
We consider a MC whose transition probability matrix is $P$, and $s$, $u$ and $d$ three distinct states such that $\alpha_{sud} = p(T_{sd} \in \mathcal{T}_{sd}^{u}) < 1$. Let $\bar{P}$ be the transition matrix of the same MC but where both states $u$ and $d$ are made absorbing, and whose entries are thus
\begin{equation}
\label{eq:P_bar}
\bar{P}_{ij} = 
\begin{cases}
0 & \text{if } i= u,d \text{ and } i \neq j, \\
1 & \text{if } i = u,d \text{ and } i = j,\\
P_{ij} &  \text{otherwise}.
\end{cases}
\end{equation}
Next, we define a second matrix $P'$, obtained by a transformation of the matrix $\bar{P}$
\begin{equation}
\label{eq:P_prime}
P'_{ij} = 
\begin{cases}
\frac{1 - \alpha_{jud}}{1 - \alpha_{iud}} \bar{P}_{ij} & \text{if } \alpha_{iud}\neq 1,\\
\bar{P}_{ij} &  \text{otherwise}.
\end{cases}
\end{equation}
\vspace*{0.25cm}
\begin{lemma}
\label{thm:trajCondEntropy}

$\left(\text{i}\right)$~The matrix $P'$ is stochastic and $\left(\text{ii}\right)$~If $T'_{sd}$ is a random trajectory defined on the MC whose 
transition probability matrix is $P'$ then
\begin{equation*}
H(T_{sd}| T_{sd} \notin \mathcal{T}_{sd}^{u}) = H(T'_{sd}).
\end{equation*}  
\end{lemma}
\vspace*{0.25cm}
\begin{proof}
$\left(\text{i}\right)$~The matrix $\bar{P}$ is the transition probability matrix of a MC where the states $u$ and $d$ 
are absorbing. We can therefore introduce the vectors of absorption probability 
$\vect{a_{u}} = (a_{1u}, a_{2u}, \ldots, a_{nu})$ and $\vect{a_{d}} = (a_{1d}, a_{2d}, \ldots, a_{nd})$ 
where $a_{iu}$ and $a_{id}$ are, respectively, the probability of being absorbed by $u$ and $d$, given that the 
trajectory starts at $i$. These vectors are eigenvectors of $\bar{P}$ associated with the unit 
eigenvalue~\cite[p. 227]{Stewart:PMCQS} 
\begin{equation}
\label{eq:eigenvectors}
\bar{P} \vect{a_{u}} = \vect{a_{u}} \qquad \bar{P} \vect{a_{d}} = \vect{a_{d}}.
\end{equation}
Moreover as ${MC}_{\bar{P}}$ has only two absorbing states $u$ and $d$, for all $i$, $a_{iu} = 1 - a_{id}$. 
Recall that for all $i$, $\alpha_{iud}= a_{iu}$ hence $\eqref{eq:P_prime}$ can be written as 
\begin{equation*}
P'_{ij} = 
\begin{cases}
\frac{a_{jd}}{a_{id}} \bar{P}_{ij} & \text{if } a_{id} \neq 0,\\
\bar{P}_{ij} &  \text{otherwise}.
\end{cases}
\end{equation*}
Note that all transitions leading to state $u$ in $MC_{\bar{P}}$ will have zero probability in $MC_{P'}$. 
In fact, consider a state $i$ such that $\bar{P}_{iu}>0$ and $a_{id} > 0$. In the new matrix $P'$,
the probability of transition from $i$ to $u$ will be $P'_{iu} = a_{ud}\bar{P}_{iu}/a_{id}$ , 
which is zero because $a_{ud} = 0$. 
Proving that $P'$ is stochastic is now straightforward: First, the entries 
of $P'$ are positive; Second, they are properly normalized and sum up to one. Indeed, if we consider a state $i$ such that 
$a_{id} = 0$, we have
that $\sum_{j} P'_{ij} = \sum_{j} \bar{P}_{ij} = 1$ whereas if $a_{id} \neq 0$, we have that 
\begin{align*}
\sum_{j} P'_{ij} &= \sum_{j} \dfrac{a_{jd}}{a_{id}} \bar{P}_{ij} \\ 
&= \frac{1}{a_{id}} \sum_{j} \bar{P}_{ij} a_{jd} \\ 
& = \frac{1}{a_{id}} (\bar{P} \vect{a_d})_{i}=\frac{1}{a_{id}} a_{id}=1
\end{align*} because of ($\ref{eq:eigenvectors}$).

$\left(\text{ii}\right)$~Let $p$ and $p'$ be the probability measures defined, respectively, for $MC_{P}$ and $MC_{P'}$ on 
the same sample space $\mathcal{T}_{sd}$. 
Any trajectory from the set $\mathcal{T}_{sd}$ has the form $t_{sd} =sx_2...x_kd$. 

If ${t}_{sd} \in \mathcal{T}_{sd}^{u}$, \begin{equation} p'({t}_{sd}) = 0 \label{eqn:condProb1} \end{equation} since we have constructed $MC_{P'}$ such that all transitions leading to state $u$ have zero probability. 

If ${t}_{sd} \notin \mathcal{T}_{sd}^{u}$, we have 
\begin{align}
p'({t}_{sd}) &=  P'_{sx_2}P'_{x_2x_3} \ldots P'_{x_kd} \nonumber \\   
& =  \frac{a_{x_2d}}{a_{sd}} \bar{P}_{sx_2} \frac{a_{x_3d}}{a_{x_2d}} \bar{P}_{x_2x_3}  \ldots \frac{a_{dd}}{a_{x_kd}} \bar{P}_{x_kd} \nonumber \\
& =  \frac{a_{dd}}{a_{sd}} \bar{P}_{sx_2}  \bar{P}_{x_2x_3}  \ldots \bar{P}_{x_kd}, \label{eq:TrajCondProb}
\end{align}
but $a_{dd}=1$ as the probability to be absorbed by state $d$, given that we have started at this same 
state, is $1$. Moreover, we know from~\eqref{eq:P_bar} that $P_{ij}=\bar{P}_{ij}$, for all $i \neq u, d$. As we have 
supposed that the trajectory ${t}_{sd}$  does not admit either $u$ or $d$ as intermediate states, 
$\bar{P}_{sx_2}  \bar{P}_{x_2x_3}  \ldots \bar{P}_{x_kd} = P_{sx_2}P_{x_2x_3} \ldots P_{x_kd}$. 
Rewriting (\ref{eq:TrajCondProb}) yields
\begin{align}
p'({t}_{sd}) &= \frac{1}{a_{sd}} P_{sx_2}P_{x_2x_3} \ldots P_{x_kd} \nonumber \\ 
& = \frac{p({t}_{sd})}{1 - a_{su}} \nonumber \\ 
&=  \frac{p({t}_{sd})}{1- p(T_{sd} \in \mathcal{T}_{sd}^{u})} = p({t}_{sd}|{T}_{sd} \notin \mathcal{T}_{sd}^{u}).\label{eqn:condProb2} 
\end{align}
Combining~\eqref{eqn:condProb1} and~\eqref{eqn:condProb2}, we have therefore proven, for all ${t}_{sd} \in \mathcal{T}_{sd}$, that 
\begin{equation}
\label{eq:entropy_equality}
p'(t_{sd}) = p({t}_{sd}|{T}_{sd} \notin \mathcal{T}_{sd}^{u}).
\end{equation}
Consequently, if the random variable $T'_{sd}$ describes the trajectory between $s$ and $d$ in $MC_{P'}$, \eqref{eq:entropy_equality} implies that
\begin{equation*}
H({T}_{sd}| {T}_{sd} \notin \mathcal{T}_{sd}^{u}) = H(T'_{sd}).
\end{equation*}\end{proof}
For the particular case where $s=d$, we still can use Lemma~\ref{thm:trajCondEntropy} to express the conditional entropy $H_{ss \vert \bar{u}}$: We modify the MC by removing the incoming transitions of $s$ and creating a new state $s'$ that will inherit them. The conditional entropy $H_{ss \vert \bar{u}}$ in the original MC is equal to $H_{ss' \vert \bar{u}}$ in the modified one and, since $s \neq s'$, we can use Lemma~\ref{thm:trajCondEntropy} to express it.

Building on Lemma~\ref{thm:cond_entropy_as_sum} and Lemma~\ref{thm:trajCondEntropy}, we can now state the main result of this paper: a general expression for the entropy of Markov trajectories conditional on multiple intermediate states.

\vspace*{0.25cm}
\begin{theorem}
\label{thm:main_result}
Let $P$ be the transition probability matrix of a finite Markov chain and $s \vect{u} d = su_1 \ldots u_ld$ a sequence of states such that $p(T_{sd} \in \mathcal{T}_{sd}^{\vect{u}}) > 0$. Then, we have the following equality 
\begin{equation}
H(T_{sd}\vert T_{sd} \in \mathcal{T}_{sd}^{\vect{u}}) = \sum_{k=0}^{l-1} H(T'_{u_ku_{k+1}}) + H(T_{u_ld}),
\end{equation}
where $u_0 = s$, and $T'_{u_ku_{k+1}}$ is a random trajectory defined on the Markov chain whose transition probability matrix $P'_{k}$  is defined as follows 
\begin{equation}
\label{eq:MC_transformation}
(P'_{k})_{ij} = 
\begin{cases}
0 & \text{if } i= u_{k+1},d \text{ and } i \neq j, \\
1 & \text{if } i= u_{k+1},d \text{ and } i = j, \\
P_{ij} & \text{if } i \neq u_{k+1},d \text{ and } \alpha_{idu_{k+1}} = 1, \\
\frac{1 - \alpha_{jdu_{k+1}}}{1 - \alpha_{idu_{k+1}}} P_{ij} & \text{if } i \neq u_{k+1},d \text{ and } \alpha_{idu_{k+1}} < 1.
\end{cases}
\end{equation} 
\end{theorem}
\vspace*{0.25cm}

\begin{proof}
The matrix $P'_{k}$ is obtained from $P$ using \eqref{eq:MC_transformation}, which is equivalent to applying successively \eqref{eq:P_bar} and \eqref{eq:P_prime} where the starting, intermediate and ending states are, respectively, $u_k$, $d$ and $u_{k+1}$. Therefore, using Lemma~\ref{thm:trajCondEntropy}, we have 
\begin{equation*}
H(T'_{u_ku_{k+1}}) = H({T}_{u_ku_{k+1}}| {T}_{u_ku_{k+1}} \notin \mathcal{T}_{u_ku_{k+1}}^{d})
\end{equation*}
for all $0 \leq k \leq l-1$. Consequently, we can write that 
\begin{align*}
& \sum_{k=0}^{l-1} H(T'_{u_ku_{k+1}}) + H(T_{u_ld}) \\
&= \sum_{k=0}^{l-1} H({T}_{u_ku_{k+1}}|{T}_{u_ku_{k+1}} \notin \mathcal{T}_{u_ku_{k+1}}^{d}) + H(T_{u_ld}),
\end{align*} where $u_0 = s$.
Using Lemma~\ref{thm:cond_entropy_as_sum}, we finally obtain
\begin{equation*}
\sum_{k=0}^{l-1} H(T'_{u_ku_{k+1}}) + H(T_{u_ld}) = H(T_{sd}\vert T_{sd} \in \mathcal{T}_{sd}^{\vect{u}}).
\end{equation*}
\end{proof}

Now that we have derived a general expression for the entropy of Markov trajectories conditional on multiple states, we introduce, in the next section, a method that allows us to compute this expression.  
\section{Entropy Computation}
The closed-form expression for the entropy of Markov trajectories proposed by Ekroot and Cover~\cite{EntropMarkTraj} is valid only if the Markov chain studied is irreducible. However, the Markov chain $MC_{P'}$ obtained from $MC_{P}$ after the transformations~\eqref{eq:P_bar} and \eqref{eq:P_prime} is not necessarily irreducible: all transitions leading to state $u$ have zero probability, which implies that possibly many states do not admit any path leading to $d$.  
Therefore, we need an expression for the entropy of Markov trajectories that is valid under milder conditions. In order to identify these conditions, we study the properties of $MC_{P'}$.  
Let $\mathcal{S}$ be the set of all states in $MC_{P'}$ and let $\mathcal{S}_1$ and $\mathcal{S}_2$ be two subsets that 
partition $\mathcal{S}$ in the following manner 
\begin{equation*}
\mathcal{S}_1 = \lbrace i \in \mathcal{S}: a_{id} > 0 \rbrace  \qquad \mathcal{S}_2 = \lbrace i \in \mathcal{S}: a_{id} = 0 \rbrace.
\end{equation*}
The set $\mathcal{S}_1$ is closed as no one-step transition is possible from any state in $\mathcal{S}_1$ 
to any state in $\mathcal{S}_2$. In fact, if $i \in \mathcal{S}_1$ and $j \in \mathcal{S}_2$, $\eqref{eq:P_prime}$ yields that
$P'_{ij} =  \bar{P}_{ij} a_{jd}/a_{id} = 0.$
Clearly, all trajectories leading to state $d$ are composed of states belonging to $\mathcal{S}_1$. 
Now, we propose a closed-form expression for the entropy of Markov trajectories that is valid under 
the weaker condition that the destination state $d$ can be reached from any other state of the MC.
Moreover, we prove that the trajectory entropy can be expressed as a weighted 
sum of local entropies.  We also provide an intuitive interpretation of the weights.

\vspace*{0.25cm}
\begin{lemma}
\label{thm:trajEntropy2}
Let $P$ be the transition probability matrix of a finite state MC such that
there exists a path with positive probability from any state to a given state $d$. Let $Q_d$ be a 
sub-matrix of $P$ obtained by removing the $d^{\text{th}}$ row and column of 
$P$.
\begin{equation*}
P = 
\left(
\begin{array}{cc|c}
\multicolumn{2}{c}{\multirow{2}{*}{\Huge{Q}\large d}} & P_{1d} \\ [-4pt]
& & \vdots \\ \hline
P_{d1} & \cdots & P_{dd} \\
\end{array}.
\right)
\end{equation*}
For any state $s \neq d$, the trajectory entropy $H_{sd}$ can be expressed as
\begin{equation}
\label{eq:entropy_fundamental}
H_{sd} = \sum_{k \neq d} ((I-Q_d)^{-1})_{sk} H(P_{k \cdot}), 
\end{equation}
where $H(P_{k \cdot})$ is the local entropy of state $k$.
\end{lemma}
\vspace*{0.25cm}

\begin{proof}
First, observe that the matrix $Q_d$ is a sub-matrix of $P$ corresponding to all 
states except state $d$ and that we use $Q_d$ to derive the entropy of all trajectories ending at $d$. 
Applying the chain rule for entropy, we express the entropy of a trajectory as the 
entropy of the first step plus the conditional entropy of the rest of the trajectory 
given this first step
\begin{equation*}
H_{sd} = H(P_{s \cdot}) + \sum_{k \neq d} P_{sk} H_{kd}.
\end{equation*}
We expand this equality further by recursively expanding the entropy $H_{kd}$ as follows 
\begin{align}
H_{sd} & =  H(P_{s \cdot}) + \sum_{k \neq d} P_{sk} \left(H(P_{k \cdot}) + \sum_{k' \neq d} P_{kk'} H_{k'd}\right) \nonumber \\
& =  H(P_{s \cdot}) + \sum_{k \neq d} P_{sk} H(P_{k \cdot}) \nonumber \\
&\quad + \sum_{k \neq d} P_{sk} \sum_{k' \neq d} P_{kk'} H_{k'd} \nonumber \\ 
& =  H(P_{s \cdot}) + \sum_{k \neq d} P_{sk} H(P_{k \cdot}) + \sum_{k \neq d} P_{sk} \sum_{k' \neq d} P_{kk'}  \nonumber \\ 
& \quad  \cdot \Bigg( H(P_{k' \cdot}) + \sum_{k'' \neq d} P_{k'k''} \Bigg( H(P_{k'' \cdot})+ \dotso \Bigg) \Bigg) \nonumber \\
& =  H(P_{s \cdot}) + \sum_{k \neq d} \left(\sum_{i=1}^{\infty} ({Q_d}^i)_{sk}\right) H(P_{k \cdot}) \nonumber \\
& =  \sum_{k \neq d} \left(\sum_{i=0}^{\infty} ({Q_d}^i)_{sk}\right) H(P_{k \cdot}), \label{eq:locEntropy1}
\end{align} with ${Q_d}^0 = I$.

Observe that the matrix $Q_d$ describes the Markov chain as long as it does not reach state $d$. Moreover, the matrix 
$Q_d$ has a finite number of states and there is a path with positive probability from each state to state $d$. As a consequence, the Markov process
will enter state $d$ with probability $1$, i.e., $\lim_{n \rightarrow \infty} {Q_d}^n = O$ (zero matrix). In addition, since 
$$(I - Q_d) (I + Q_d + {Q_d}^2 + \ldots + {Q_d}^{n-1}) = I - {Q_d}^n,$$ 
we can easily verify that 
\begin{equation}
\sum_{i = 0}^{\infty} {Q_d}^i = (I-Q_d)^{-1}. \label{eq:locEntropy2}
\end{equation}
Replacing (\ref{eq:locEntropy2}) in (\ref{eq:locEntropy1}), we have 
\begin{equation*}
H_{sd} = \sum_{k \neq d} ((I-Q_d)^{-1})_{sk} H(P_{k \cdot}).\end{equation*}\end{proof}
We have shown that the entropy of a family of trajectories can be expressed as a weighted sum of 
the states' local entropies. The weights are given by the matrix $(I-Q_d)^{-1}$.  
In the Markovian literature, the matrix $(I-Q_d)^{-1}$ is referred to as the fundamental matrix ~\cite{Stewart:PMCQS,snell:FiniteMC}. In fact, the $(sk)^{\text{th}}$ element of the fundamental matrix  
(defined with respect to the destination state $d$) can be seen as the expected number of visits to the state $k$ before hitting the state $d$, given that we started at state $s$. As a result, the entropy of the random trajectory 
${T}_{sd}$ is the sum over the chain states of the expected number of visits to each state multiplied by 
its local entropy. This is a remarkable observation since it links a global quantity, which is the trajectory entropy, to the local entropy at each state. 

Recall that in the example shown in Figure~\ref{fig:MCexample}, we found that the entropy of the trajectory $T_{15}$ is equal to $1.56$ bits. 
We can retrieve this result by computing the fundamental matrix with respect to state $5$. The $(ij)^{\text{th}}$ element of this matrix
is equal to the expected number of visits to state $j$ before hitting state 
$5$, given that we started at state $i$. Multiplying the first row of the fundamental matrix 
$(1,0.625,0.75,0.375) $ by the column vector of local entropies $(0.81,0,1,0)$ yields 
$H_{15} =1 \times 0.81 + 0.75 \times 1 = 1.56$ bits. 

\subsection{Algorithm}
The following algorithm defines the set of steps to compute the entropy of Markov trajectories conditional on a set of intermediate states:

\begin{figure}[h]
\label{alg:cond_entr}
\begin{boxedalgorithmic}[1]
\REQUIRE Matrix of transition probability $P$, source state $s$, destination state $d$, sequence of intermediate states~$\vect{u}~=~u_1 \ldots u_l$
\ENSURE  $H_{sd\vert u_1\ldots u_l}$
\STATE $u_0 \leftarrow s$
\FOR{$k=0$ to $l-1$}
\STATE Compute $P'_k$ from $P$ using \eqref{eq:MC_transformation} 
\STATE Compute $H(T'_{u_k u_{k+1}})$ from $P'_k$ using Lemma~\ref{thm:trajEntropy2}
\STATE $H_{u_k u_{k+1}\vert \bar{d}} \leftarrow H(T'_{u_k u_{k+1}})$ \COMMENT{Lemma~\ref{thm:trajCondEntropy}}
\ENDFOR
\STATE Compute $H_{u_ld}$ from $P$ using Lemma~\ref{thm:trajEntropy2}
\STATE $H_{sd\vert u_1\ldots u_l} =  \sum_{k=0}^{l-1} H_{u_ku_{k+1} \vert \bar{d}}+ H_{u_ld}$ \COMMENT{Lemma~\ref{thm:cond_entropy_as_sum}}
\RETURN $H_{sd\vert u_1\ldots u_l}$
\end{boxedalgorithmic}
\end{figure}

The worst-case running time for the algorithm is $\mathcal{O}(lN^3)$ where $N$ is the number of states of $MC_{P}$, and $l$ the length of the sequence of intermediate states $\vect{u}$. This complexity is dominated by the cost of computing the inverse of the matrix $(I - Q_d)$, which is needed to compute the entropy $H_{sd}$ in~\eqref{eq:entropy_fundamental}. However, since we need only the $s^{\text{th}}$ row of the matrix $(I - Q_d)$ to compute the trajectory entropy $H_{sd}$, we can solve a system of\textemdash potentially sparse\textemdash linear equations. Moreover, many iterative methods~\cite[p.~508]{golub:matrix} take advantage of the structure of the matrix representing the system of linear equations in order to solve them efficiently.

Coming back to the example shown in Figure~\ref{fig:MCexample}, we use the algorithm above to compute the conditional entropy $H_{15 \vert 3} = 1$ bit. We leave no ambiguity about the trajectory $T_{15}$ when we condition on both states $3$ and $2$ and find that $H_{15 \vert 3,2} = H_{13 \vert \bar{5}} + H_{32 \vert \bar{5}} + H_{25} = 0$ bits.

\paragraph*{Conditioning on a set of states}
\begin{figure}[hdt]
\centering
\epsfig{file=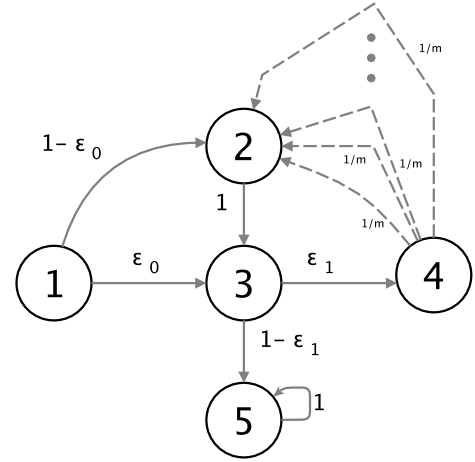,height=2.5in, width=2.5in}
\caption{A Markov chain annotated with the transition probabilities. The 
dashed lines between states 4 and 2 represent the $m$ equiprobable paths leading 
from state 4 to state 2. We choose $0 < \epsilon_1 < 1$ and $m \geq 1$ to guarantee that 
$|\mathcal{T}_{15}| > 0$ and that $p ( T_{15} \in \mathcal{T}_{15}^{3,2}) >0$.}
\label{fig:MCexample_2}
\end{figure}
In this paper, we focused on computing the entropy of Markov trajectories conditional 
on a \emph{sequence} of states. A natural extension is the computation of this entropy conditional
on a \emph{non ordered} set of states. Finding a general expression for this 
conditional entropy appears very hard and there is no simple relation linking it to the entropy conditional 
on a sequence. We provide an example, shown in Figure~\ref{fig:MCexample_2}, 
that illustrates an interesting and counter-intuitive result about conditioning on a set of states.
Intuitively, we would expect that the entropy of a random trajectory conditional on 
a sequence of states is always less than the entropy of the same trajectory conditional on the 
set formed by these states. However, this is not true. 
We take the MC shown in Figure~\ref{fig:MCexample_2} as an example and we compute, using 
Theorem~\ref{thm:main_result}, the entropy of the random trajectory $T_{15}$
conditional on going through the sequence of intermediate states $\left( 3,2 \right)$  
\begin{align}
\label{eq:example_entropy}
H_{15|32} &= H_{13|\bar{5}} + H_{32|\bar{5}} + H_{25} \nonumber \\
& = h(\epsilon_0) + \log m + H_{35},
\end{align}
where $h(\epsilon_0)$ is the entropy of a Bernoulli random variable with success probability $\epsilon_0$. 
To compute the entropy of the random trajectory $T_{15}$ conditional on going through the 
set of states $\lbrace 2,3 \rbrace$, we apply the chain rule for entropy and express the entropy of a trajectory 
as the entropy of the first two steps plus the conditional entropy of the rest of the trajectory 
given these first two steps
\begin{align*}
H_{15|\lbrace 2,3\rbrace} & = h \left(\frac{\epsilon_0 \epsilon_1}{1 - \epsilon_0(1-\epsilon_1)} \right) + 
\frac{\epsilon_0 \epsilon_1}{1 - \epsilon_0(1-\epsilon_1)} H_{45} \\
&\quad + \frac{1-\epsilon_0}{1 - \epsilon_0(1-\epsilon_1)} H_{35}. 
\end{align*}
Since $H_{45} = \log m + H_{25} = \log m + H_{35}$, we have that
\begin{align}
H_{15|\lbrace 2,3\rbrace} &= h \left(\frac{\epsilon_0 \epsilon_1}{1 - \epsilon_0(1-\epsilon_1)} \right) + 
\frac{\epsilon_0 \epsilon_1}{1 - \epsilon_0(1-\epsilon_1)} \log(m) \nonumber \\
&\quad + H_{35} \label{eq:example_entropy_2}.
\end{align}
Using \eqref{eq:example_entropy} and \eqref{eq:example_entropy_2}, we can write
\begin{align*}
H_{15|32} - H_{15|\lbrace 2,3\rbrace} &=  h(\epsilon_0) - h \left(\frac{\epsilon_0 \epsilon_1}{1 - \epsilon_0(1-\epsilon_1)} \right) \\
&\quad + \frac{1-\epsilon_0}{1 - \epsilon_0(1-\epsilon_1)} \log m.
\end{align*}
This difference can therefore be lower bounded by 
\begin{align*}
H_{15|32} - H_{15|\lbrace 2,3\rbrace} \geq -1 +  \dfrac{1-\epsilon_0}{1 - \epsilon_0(1-\epsilon_1)} \log m.
\end{align*}
As a consequence, if $\log m > 1 + {\epsilon_0 \epsilon_1}/{1-\epsilon_0}$, the entropy of the random 
trajectory $T_{15}$ conditional on going through the sequence $\left( 3,2 \right)$ is strictly greater 
than the entropy of the same trajectory conditional on going through the set of states $\lbrace 2,3 \rbrace$.
The reason is that conditioning on the sequence $\left( 3,2 \right)$ implies that the random trajectory $T_{15}$ is 
composed of a random sub-trajectory $T_{42}$ whose entropy can be made arbitrary large by 
increasing the parameter $m$. More generally, this example illustrates the absence of a simple 
relation between the entropy of random trajectories conditional on a sequence of states and the entropy of the same 
trajectory conditional on the set formed by these same states.      

\section{Conclusion}
In this paper, we address the problem of computing the entropy of conditional Markov trajectories. 
We propose a method based on a transformation of the original Markov chain into a Markov chain 
that yields the desired conditional entropy. We also derive an expression that allows us to compute the entropy of Markov trajectories, under conditions weaker than those assumed in~\cite{EntropMarkTraj}. Furthermore, this expression links the entropy of Markov trajectories\textemdash a global quantity\textemdash to the local entropy of states.  

These results have applications in various fields including mobility privacy of the users of online services. 
In fact, using our framework, we are able to quantify the predictability of a user's mobility and its 
evolution with locations updates: We represent a location as a state of a Markov chain. A sequence of visited locations
is therefore a Markovian trajectory, and location-updates amount to conditioning this trajectory on a set of intermediate states. 
In this setting, we can quantify the evolution of the user's mobility predictability as she/he discloses some of the locations she/he visited by computing the entropy of conditional Markov trajectories. 
Consequently, users are empowered with an objective technique to protect their privacy: they are able to anticipate the evolution of their mobility predictability as they reveal a subset of the locations they visited. 

\section*{Acknowledgment}
The authors would like to thank Olivier L\'{e}v\^{e}que and Emre Telatar for their
feedback about this paper.

\begin{IEEEbiographynophoto}{Mohamed Kafsi}(S'12) is a Ph.D. student in the School of Computer and Communication Sciences at EPFL. His research 
interests lie at the intersection of graph theory, information theory and data mining. He received the M.S. and B.S. degrees in 
communication systems from EPFL. During his Bachelor's studies, he spent one year at the Electrical and Computer Engineering  department of 
Carnegie Mellon University (CMU). He is a member of the winning team of the 2013 Nokia Mobile Data Challenge. 
\end{IEEEbiographynophoto}

\begin{IEEEbiographynophoto}{Matthias Grossglauser}(M'95)
is an Associate Professor in the School of Computer and Communication Sciences at EPFL. 
He received his Dipl\^{o}me d'Ing\'{e}nieur en Syst\`{e}mes de Communication degree from EPFL in 1994, the M.Sc. degree from the Georgia Institute of Technology in 1994, and the Ph.D. from the University Pierre et Marie Curie (Paris 6) in 1998. 
His research interests are in social and information networks, privacy, mobile and wireless networking, and network traffic measurement and modeling. He received the 1998 Cor Baayen Award from the European Research Consortium for Informatics and Mathematics (ERCIM), the IEEE INFOCOM 2001 Best Paper Award, and the 2006 CoNEXT/SIGCOMM Rising Star Award. 
He served on the editorial board of IEEE/ACM Transactions on Networking, and on numerous Technical Program Committees.
From 2007-2010, he was with the Nokia Research Center (NRC) in Helsinki, Finland, holding the positions of Laboratory Director, then of Head of a tech-transfer program focused on data mining, analytics, and machine learning. 
In addition, he served on Nokia's CEO Technology Council, a technology advisory group reporting to the CEO. 
From 2003-2007, he was an Assistant Professor at EPFL. 
From 1998 to 2002, he was a Senior, then Principal Member of Research Staff in the Networking and Distributed Systems Laboratory at AT\&T Research in New Jersey. From 1995 to 1998, he was a Ph.D. student at INRIA Sophia Antipolis, France.
\end{IEEEbiographynophoto}

\begin{IEEEbiographynophoto}{Patrick Thiran}(S'89 -- M'96 -- SM'12)
received the electrical engineering degree from
the Universit\'{e} Catholique de Louvain, Louvain-la-Neuve, Belgium, in 1989,
the M.S. degree in electrical engineering from the University of California at Berkeley, USA,
in 1990, and the Ph.D. degree from EPFL, in 1996.
He is a Full Professor at EPFL. He became an Adjunct Professor in
1998, an Assistant Professor in 2002, an Associate Professor in 2006
and a Full Professor in 2011. From
2000 to 2001, he was with Sprint Advanced Technology Labs, Burlingame, CA.
His research interests include communication networks, performance analysis,
dynamical systems, and stochastic models. He is currently active in the analysis
and design of wireless and PLC networks, in network monitoring, and in data-driven
network science.
Dr. Thiran served as an Associate Editor for the IEEE Transactions on
Circuits and Systems in 1997-99, and as an Associate Editor
for the IEEE/ACM Transactions on Networking in 2006-10. He was the recipient of
the 1996 EPFL Ph.D. award and of the 2008 Cr\'{e}dit Suisse Teaching Award.
\end{IEEEbiographynophoto}

\end{document}